\documentclass[%
 reprint,
 superscriptaddress,
 amsmath,amssymb,
 aps,
prb,
floatfix,
]{revtex4-2}

\usepackage{graphicx}
\usepackage{dcolumn}
\usepackage{bm}

\newcommand{\widePicWidth}{17.5cm}

\usepackage{hyperref}
\hypersetup{colorlinks,allcolors=blue}

\begin{document}

\title{Improving Spectral Resolution from Real-time Evolution for Correlated Systems}

\author{Ta Tang}
\email{tatang@stanford.edu}
\affiliation{Stanford Institute for Materials and Energy Sciences,
SLAC National Accelerator Laboratory, 2575 Sand Hill Road, Menlo Park, CA 94025, USA}
\affiliation{Department of Applied Physics, Stanford University, Stanford, CA 94305, USA}

\author{Chunjing Jia}
\affiliation{Department of Physics and Quantum Theory Project, University of Florida, 2001 Museum Road, Gainesville, 32611, Florida, United States}

\author{Brian Moritz}
\affiliation{Stanford Institute for Materials and Energy Sciences,
SLAC National Accelerator Laboratory, 2575 Sand Hill Road, Menlo Park, CA 94025, USA}

\author{Thomas P. Devereaux}
\email{tpd@stanford.edu}
\affiliation{Stanford Institute for Materials and Energy Sciences,
SLAC National Accelerator Laboratory, 2575 Sand Hill Road, Menlo Park, CA 94025, USA}
\affiliation{Department of Materials Science and Engineering, Stanford University, Stanford, CA 94305, USA}
\affiliation{Geballe Laboratory for Advanced Materials, Stanford University, Stanford, California 94305, USA}

\date{\today}
\begin{abstract}
    The quality of numerically simulated spectra using real-time evolution methods for strongly correlated systems is affected by both the length of simulation time and the system size, limiting resolution in both frequency and momentum. 
    In this work, we propose a computationally cheap, linear autoregressive machine learning-based framework to extend short-time and distance results over a wider range. We demonstrate the proposed method to extend the lesser Green's function for both the Hubbard model and the much more computationally challenging Hubbard-extended Holstein model. This technique significantly improves both the frequency and momentum resolution of the single-particle removal spectrum $\mathcal{A}(k,\omega)$, allowing observation of otherwise obscured spectral features due to electron-phonon coupling.
\end{abstract}
\maketitle

\section{Introduction}

Machine learning has emerged as a promising tool in condensed matter and materials physics, enabling new insights across a broad range of quantum many-body problems. Among the most pressing challenges in this field is the development of numerical algorithms capable of tackling strongly correlated systems at larger system sizes, in higher dimensions, and over longer time scales without sacrificing the accurate treatment of long-range entanglement and strong correlation effects. In recent years, neural-network quantum states (NQS) have been introduced as a variational ansatz for many-body wavefunctions, offering a route to simulate model Hamiltonians such as the Hubbard and Heisenberg models \cite{carleo2017, Sharir2020}. These wavefunction-based approaches have the potential to move beyond the quasi-one-dimensional regime traditionally accessible to density matrix renormalization group (DMRG) methods; however, the growth of entanglement entropy of the wave functions~\cite{vidalEntanglementQuantum2003, vidalEfficientClassical2003,vidalEfficientSimulation2004} leads to a growing computational cost with longer times and larger system sizes.

In this work, we propose a different type of novel machine learning-based framework that utilizes a linear autoregressive model to directly learn from spectral functions, or more precisely correlation functions $\chi(r,t)$, which may have a reduced complexity and present certain patterns that could be utilized to make predictions about values at later times and longer distances. This bypasses the need to represent the many-body wavefunction itself and accurately account for the growing entanglement entropy. The framework offers a new pathway to reach larger length and time scales that enable the study of dynamical correlations in strongly correlated systems with unprecedented efficiency and scalability. By leveraging machine learning for strongly correlated systems in a novel way, our approach opens the door to addressing longstanding challenges in quantum magnetism and high-temperature superconductivity that have remained beyond the reach of conventional numerical techniques.

Dynamical quantities $\chi(q,\omega)$ are essential for characterizing excitations in strongly correlated systems and can be used as a bridge between results from modeling and those from experimental measurements. $\chi(q,\omega)$ can be computed via its Lehmann representation, for example by using the exact diagonalization (ED) method with Lanczos iterations to obtain dynamical quantities~\cite{dagottoCorrelatedElectrons1994}. While ED is unbiased and versatile, it is limited to very small cluster sizes. 
A more powerful method for one-dimensional (1D), or quasi-1D systems, is density matrix renormalization group (DMRG)~\cite{whiteDensityMatrix1992, whiteDensitymatrixAlgorithms1993}. 
Ground state and finite-temperature approaches~\cite{PhysRevLett.102.190601} based on DMRG can be easily applied to very large systems ($\sim$ a few hundred sites). Extensions such as the correction vector method~\cite{PhysRevB.60.335} and variational approaches~\cite{jeckelmannDynamicalDensitymatrix2002} have been developed to successfully capture dynamical properties, denoted as dynamical DMRG (DDMRG) methods, for much larger systems compared to ED.  
However, DDMRG requires separate converged simulations for each point in $(q,\omega)$-space and can be computationally inefficient if the goal is to obtain the entire spectra.

An alternative method to calculate $\chi(q,\omega)$ is to first compute the time-dependent correlator $$\chi(r,t) = \left<\hat{O}^\dagger_{r_0+r}(t_0+t)\hat{O}_{r_0}(t_0)\right>$$ and then apply Fourier transforms. Time-dependent DMRG (tDMRG) \cite{vidalEfficientClassical2003,whiteRealTimeEvolution2004, paeckelTimeevolutionMethods2019} is an extension of DMRG, making real-time evolution possible. This leads to a very efficient method to evaluate $\chi(q,\omega)$, where the full spectra can be obtained by a single tDMRG simulation. However, computational complexity grows rapidly as the length of time and cluster size increase, leading to low-frequency and -momentum resolution, especially for more numerically challenging problems, {\it e.g.} those involving multiple degrees of freedom or extended interactions.


Previous works based on time series predictions~\cite{whiteSpectralFunction2008, tianMatrixProduct2021} to improve the resolution have only extended the correlator near the center of the system along the time direction. Here we propose a method to extend the time-dependent correlator along both the time and space directions based on the pattern of short-time and short-distance results on a small cluster, simultaneously improving both the frequency and momentum resolution of the spectra. We first demonstrate this method on the single-particle removal spectra for the 1D Hubbard model. We then turn to a much more computationally expensive problem -- the single-particle removal spectra of the 1D Hubbard-extended Holstein model with extended electron-phonon ({\it el-ph}) coupling.
Our method provides high resolution for the simulated spectra in both cases, and highlights phonon-induced spectral features that are not apparent in the raw simulated spectra because of the low resolution. 

\begin{figure*}[htp!]
    \centering
    \includegraphics[width=\widePicWidth]{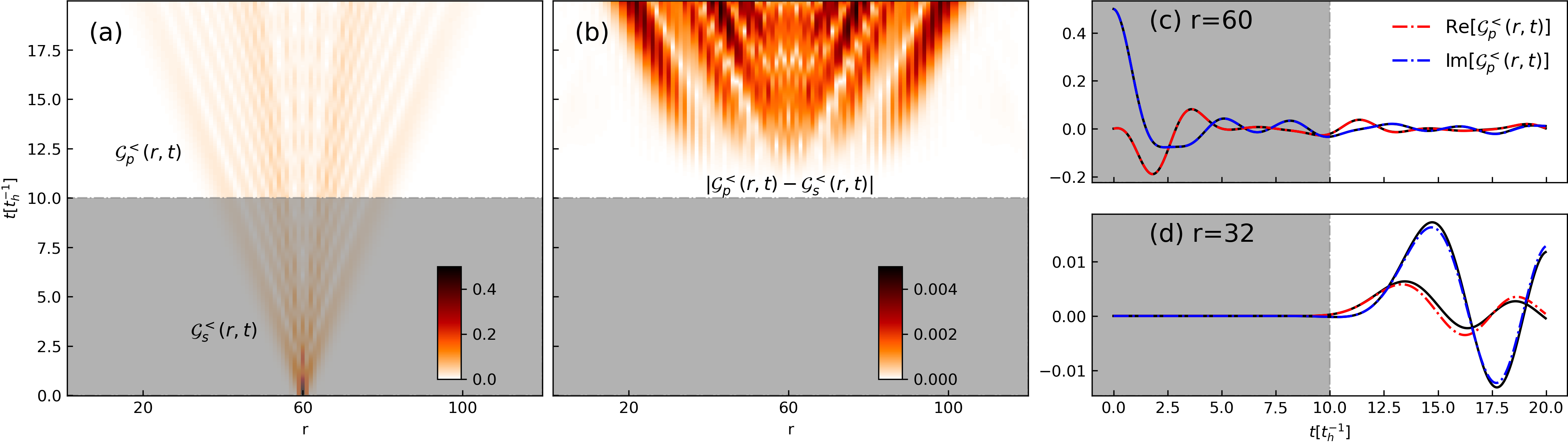}
    \caption{Comparison between predicted and simulated lesser Green's function. The simulated data is obtained on a $L=120$ half-filled Hubbard chain using tDMRG. The total time simulated is $20[t_h^{-1}]$. (a) Predicted Green's function. The first half of the simulated data $10[t_h^{-1}]$ is used for training (indicated by the gray area). The data above the gray area is predicted. (b) Difference between predicted and simulated lesser Green's function. Note that the color bar scale in (b) is 100 times smaller than that in (a). In (c) and (d), we plot the real and imaginary part of the predicted $\mathcal{G}^{<}_p(r,t)$ (red and blue dashed lines) against the simulated $\mathcal{G}^{<}_s(r,t)$ (black solid lines) at the center of the chain $r=60$ and at the boundary of the training data light cone $r=32$. The gray areas indicate data used in training. }
    \label{fig:hubbard_green_benchmark}
\end{figure*}

\section{Method}
Computationally, using a method like tDMRG is very expensive, scaling like $O(LTM^3)$, where $L$ is the system length measured in the number of sites, $T$ is the simulated time measured in time steps, and $M$ is the DMRG bond dimension, which also increases with $L$ and $T$. Previous works have tried time series predictions, where for each fixed position $r$, the future time step $\chi(r,t=n)$ is obtained by using data from previous $d_t$ time steps, {\it e.g.} using a linear autoregressive model for linear prediction (LP)~\cite{whiteSpectralFunction2008} 
\begin{equation}
    \mathcal{\chi}(r,t) = \sum_{j=1}^{d_t}\theta^{r}_j\mathcal{\chi}(r,t-j),
    \label{eq:lp}
\end{equation}
or a more sophisticated recursion method~\cite{tianMatrixProduct2021}. The process can be repeated to obtain future time step data at $t=n+2,\, n+3,\, \dots$. This can extend $\chi(r,t)$ along the time dimension, improving the frequency resolution. 
An obvious limitation of this method is that it can only extend $\mathcal{\chi}(r,t)$ along the time direction near the center portion of the chain since there is not enough data near the boundary to obtain $\theta^r$ parameters and make good predictions. Therefore, it only improves the frequency resolution but not the momentum resolution. Additionally, the accuracy may suffer since only the center portion of the data is predicted along the time direction, thus missing a large portion of the correlator data for computing the spectra.

\begin{figure*}[htp]
    \centering
    \includegraphics[width=\widePicWidth]{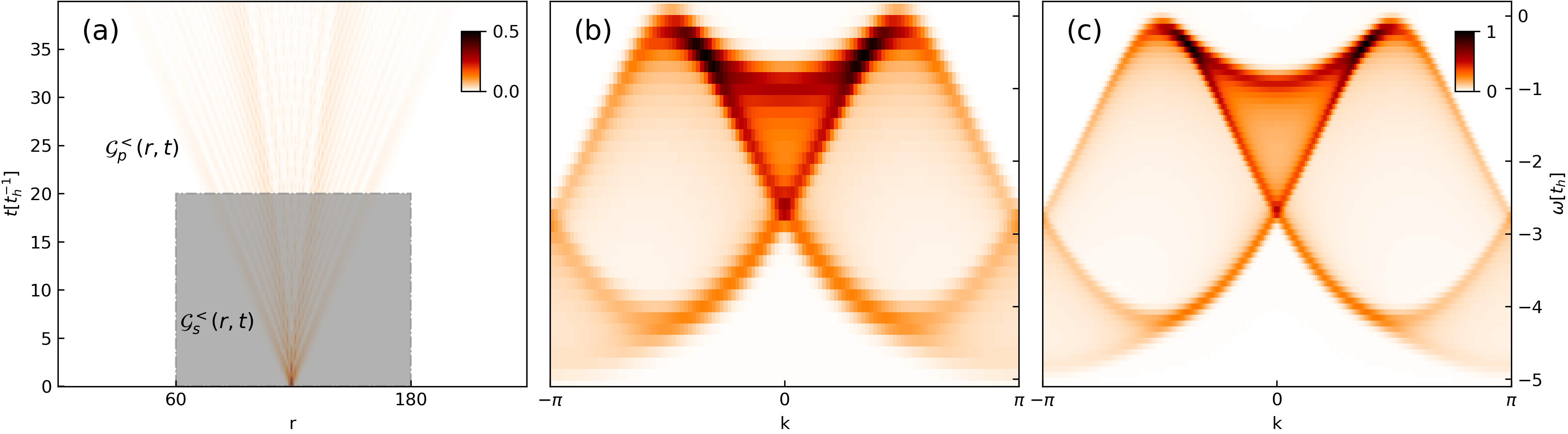}
    \caption{Extended lesser Green's function for the 1D Hubbard model. (a) Simulated and predicted lesser Green's function for the Hubbard model at half filling. The gray area indicates the simulated data obtained from tDMRG with $L_s=120$ and total simulated time $T_s = 20[t_h^{-1}]$. Using the prediction method, it is extended to $L_p=240$ and $T_p = 40[t_h^{-1}]$. (b) Single particle spectral functions obtained from tDMRG simulated lesser Green's function (gray area in (a)) with $L_s=120$ and $T_s=20[t_h^{-1}]$. The broadening used is $\sigma_{\omega}=0.16t_h$, roughly equals to the FFT frequency spacing. (c) Single particle spectral functions obtained from predicted lesser Green's function from (a) with $L_s=240$ and $T_p=40[t_h^{-1}]$. The broadening used is $\sigma_{\omega}=0.08t_h$, roughly equals to the FFT frequency spacing. The spectra in (c) has much better (2x) frequency and momentum resolutions compared to the spectra in (b).}
    \label{fig:hubbard_akw}
\end{figure*}

We propose a method to improve the shortcomings of the simple LP method by predicting the correlator data along {\it both} the time and space directions: simulations of $\mathcal{\chi}(r,t)$ on a short chain of length $L_s$ evolved it up to a short time $T_s$ are used to predict $\mathcal{\chi}(r,t)$ on a longer chain of length $L_p$ and up to a longer time $T_p$. In so doing, the resultant spectra will have a higher resolution in both frequency and momentum, as the resolution scales inversely to $T$ and $L$ ($\delta\omega\sim 2\pi/T$ and $\delta k\sim 2\pi/L$).
We make the assumption that the value of $\mathcal{\chi}$ at position $r$ and time $t$ can be determined by a linear combination of nearby $\mathcal{\chi}(r-i, t-j)$ where $-d_r \le i \le d_r$ and $1 \le j \le d_t$
\begin{equation}
    \mathcal{\chi}(r,t) \approx h_\theta(r,t) = \sum_{i=-d_r}^{d_r}\sum_{j=1}^{d_t} \theta_{ij}\mathcal{\chi}(r-i,t-j).
    \label{eq:hypothesis}
\end{equation} 
This can be formulated as a linear regression problem to find the best $\theta$ that can fit the simulated data. We use $y^m$, where $m$ maps to a 2D index $(r_m, t_m)$, to label the target of the linear prediction $ y^m = \mathcal{\chi}(r_m,t_m)$, $0\le r_m < L$ and $d_t\le t_m \le T$. We use $x^{(m)} \in R^{(2d_r+1)d_t}$ to label the input data for the target $y^m$ , and $x^{(m)}_{ij} =\mathcal{\chi}(r_m-i, t_m-j)$. The linear prediction is $h_\theta(m) = \sum_{i=-d_r}^{d_r}\sum_{j=1}^{d_t} \theta_{ij}x^{(m)}_{ij}$. We use a $L_2$ loss function
\begin{equation}
    J(\theta) = \frac{1}{2N}\sum_{m=1}^{N}|y^{(m)} - h_\theta(m)|^2,
    \label{eq:loss}
\end{equation}
where $N$ is the total number of valid examples. The optimal $\theta$ is found by minimizing this loss function. After the optimal $\theta$ is determined from the simulated data, we can then use it to extend the time-dependent correlator for longer chains ($L_p$) and later times ($T_p$) by repeatedly applying Eq.~\ref{eq:hypothesis} for $0\le r < L_p$ at $t=T_s+1,\, T_s+2,\, \dots, T_p$.

\section{Hubbard Model}
We first demonstrate the effectiveness of the approach by computing the single-particle removal spectrum of the one-dimensional Hubbard model at half-filling.
The 1D Hubbard model Hamiltonian is
\begin{equation}
    H = -t_h\sum_{\left<ij\right>\sigma}(c^\dagger_{i\sigma}c_{j\sigma} + h.c.) + U\sum_i n_{i\uparrow}n_{i\downarrow},
    \label{eq:hubbard}
\end{equation}
where $t_h$ is the nearest neighbor hopping integral, $U=8t_h$ is the on-site repulsion, $c^\dagger_{i\sigma}$ is the creation operator for spin $\sigma$ on site $i$, $c_{j\sigma}$ is the annihilation operator for spin $\sigma$ on site $j$, $n_{i\sigma}$ is the charge number on site $i$ with spin $\sigma$, $\left<ij\right>$ denotes nearest neighbor sites, and $h.c.$ represents hermitian conjugate. The single-particle removal spectra $\mathcal{A}(k, \omega)$ can be obtained by a Fourier transform of the lesser Green's function $\mathcal{G}^{<}(r,t)$ and is related to experimentally measured angle-resolved photoemission (ARPES) spectra~\cite{RevModPhys.75.473, sobotaAngleresolvedPhotoemission2021}.

We use the DMRG approach to calculate the groundstate wavefunction $\left|g\right>$, and then use tDMRG~\cite{vidalEfficientClassical2003,whiteRealTimeEvolution2004, paeckelTimeevolutionMethods2019} to compute the lesser Green's function
\begin{equation}
    \mathcal{G}^{<}(r,t) = i\left<g\right|c^\dagger_r(t)c_{r_0}(0)\left|g\right>,
    \label{eq:green}
\end{equation}
where the spin index has been omitted for brevity. $L$ is the length of the chain and $r_0=L/2-1$ or $L/2$ denotes the center of the chain. The single-particle removal spectra is obtained from a Fourier transform of the lesser Green's function
\begin{equation}
    \mathcal{A}(k,\omega) = \int^\infty_{-\infty} \frac{dt ~e^{i\omega t}}{2\pi i} \sum_{r} \frac{e^{-ik(r - r_0)}}{L} \mathcal{G}^{<}(r,t),
\end{equation}
where $k$ represents momentum along the chain and $\omega$ is the frequency. To ensure reflection symmetry for chains with an even number of sites, we average over the spectra obtained from $r_0=L/2-1$ and $L/2$. 
We evolve the system from time $0$ to a time $T$ before the excitation propagates to the boundary of the chain. To regularize the Fourier transform due to a finite cutoff in time, we use a Gaussian window function \cite{whiteRealTimeEvolution2004} $\mathcal{W}_{\sigma_{\omega}}(t) = \exp{\left[-\sigma_{\omega}^2\,t^2/2\right]}$ with a frequency domain standard deviation $\sigma_{\omega}$, and $\mathcal{W}_{\sigma_{\omega}}(T) \sim 0$, to broaden the spectra and act as a frequency resolution convolution 
$\mathcal{A}_{\sigma_{\omega}}(k,\omega) = \mathcal{A}(k,\omega) * \mathcal{F}[\mathcal{W}_{\sigma_{\omega}}(t)]$. 

\begin{figure*}[htp!]
    \centering
    \includegraphics[width=\widePicWidth]{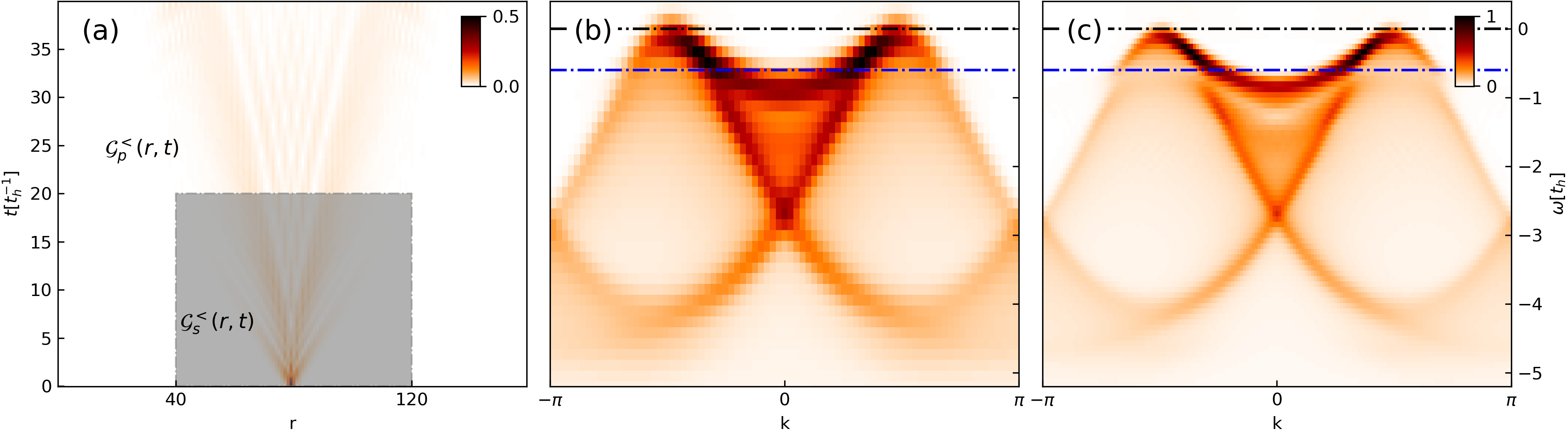}
    \caption{Extend lesser Green's function for the Hubbard-extended Holstein model. (a)The gray area indicates the simulated data obtained from tDMRG with $L_s=80$ and total simulated time $T_s = 20[t_h^{-1}]$. Using the prediction method, it is extended to $L_p=160$ and $T_p = 40[t_h^{-1}]$. (b) Hubbard-extended Holstein model single particle spectral functions at half filling obtained from tDMRG simulated lesser Green's function (gray area in (a) with $L_s=80$ and $T_s=20[t_h^{-1}]$. The broadening used is $\sigma_{\omega}=0.16t_h$, roughly equals to the FFT frequency spacing. (c) Single particle spectral functions obtained from predicted lesser Green's function from (a) with $L_p=160$ and $T_p=40[t_h^{-1}]$. The broadening used is $\sigma_{\omega}=0.08t_h$, roughly equals to the FFT frequency spacing. The spectra in (c) has much better (2x) frequency and momentum resolutions compared to the spectra in (b). The black dashed line indicates Fermi energy $E_f$ and the blue dashed line is one phonon frequency below the Fermi energy $E_f - \omega_0$. Due to the improved resolution in (c), we can clearly see replica bands separated by the phonon frequency, which is completely missing in (b).}
    \label{fig:phonon_akw}
\end{figure*}

We numerically compute $\mathcal{G}^{<}(r,t)$ on a lattice of length $L_s=120$ at half-filling and time evolve it to time $T_{s} = 20[t_h^{-1}]$. 
To demonstrate the effectiveness of our method, we use the simulated data over the first half $0\le t \le 10[t_h^{-1}]$ as the training data, and use the later half $10[t_h^{-1}] < t \le 20[t_h^{-1}]$ as the test data. We find that using $d_r=5\sim7$ and $d_t= 100\sim 150$ can produce small training and testing loss as defined in Eq.~\ref{eq:loss}.
The gray area in Fig.~\ref{fig:hubbard_green_benchmark}(a) shows simulated data $\mathcal{G}^{<}_{s}(r,t)$ used as the training data, and the data above the gray area is the predicted $\mathcal{G}^{<}_{p}(r,t)$. Fig.~\ref{fig:hubbard_green_benchmark}(b) shows the difference between the predicted and simulated lesser Green's functions. In general, the differences are small and hardly visible using the same color scale as in (a). Figure~\ref{fig:hubbard_green_benchmark}(b) uses a color scale 100 times smaller to better display the deviations, which appear at later times. The impact of these differences is reduced in the final spectra due to the Gaussian window function. We also compare the predicted and simulated lesser Green's functions at fixed locations in Fig.~\ref{fig:hubbard_green_benchmark}(c) and (d).  Even at $r=32$, near the boundary of the training data light cone where there is no historical data along the time direction, our method manages to produce decent predictions, demonstrating its advantages over previous methods. 

Applying this method to all of the simulated data, we see an immediate improvement in the spectral resolution. In Fig.\,\ref{fig:hubbard_akw}(a), we show the predicted $\mathcal{G}^{<}_p(r,t)$ with $L_p=240$ and $T_p=40[t_h^{-1}]$ based on the simulated data with $L_s=120$ and $T_s=20[t_h^{-1}]$. The patterns at longer distances and later times are consistent with those in the simulated data. In Fig.\,\ref{fig:hubbard_akw}(c), the resolution of the spectral function is greatly improved compared to that obtained from the simulated data alone, as shown in Fig.\,\ref{fig:hubbard_akw}(b), and matches well with other single-particle spectra for the Hubbard model in the literature \cite{senechalSpectralWeight2000, benthienSpectralFunction2004, kohnoSpectralProperties2010, chenAnomalouslyStrong2021, esslerOneDimensionalHubbard2005}.

\section{Hubbard-Extended Holstein Model}

With the effectiveness of our method demonstrated in the Hubbard model, we now turn to a much more computationally demanding problem involving {\it el-ph} coupling. 
Recent experimental and numerical work on doped one-dimensional cuprates has indicated that an extended attractive interaction, which likely stems from an extended {\it el-ph} interaction, is needed to explain the salient features observed in the measured ARPES data \cite{chenAnomalouslyStrong2021, wangPhononMediatedLongRange2021, tangTracesElectronphonon2023}. The numerical model used to incorporate the extended {\it el-ph} interaction and to match the measured ARPES data is the Hubbard-extended Holstein model \cite{wangPhononMediatedLongRange2021, tangTracesElectronphonon2023}
\begin{eqnarray}
    H &=& -t_h\sum_{\left<ij\right>\sigma}(\hat{c}^\dagger_{i\sigma}\hat{c}_{j\sigma} + h.c.) + U\sum_{i}\hat{n}_{i\uparrow}\hat{n}_{i\downarrow} + \omega_0\sum_i \hat{a}^\dagger_i\hat{a}_i \nonumber \\
    &&\quad + g_0\sum_i \hat{n}_i (\hat{a}^\dagger_i + \hat{a}_i) + g_1\sum_{\left<ij\right>} \hat{n}_i (\hat{a}^\dagger_j + \hat{a}_j),
    \label{eq:hhm}
\end{eqnarray}
where the first two terms are from the Hubbard model describing the electronic part, $\hat{a}^\dagger_i$ and $\hat{a}_i$ are the phonon ladder operators on site $i$, $\hat{n}_i$ is the total charge number operator on site $i$, $\omega_0$ is the phonon frequency, $g_0$ is the on-site {\it el-ph} coupling, $g_1$ is the nearest-neighbor {\it el-ph} coupling, and $\left<ij\right>$ sums over nearest-neighbors.

Although there are techniques such as local basis optimization (LBO) \cite{zhangDensityMatrix1998} and dynamical LBO \cite{brocktMatrixproductstateMethod2015} to make solving for the groundstate and performing time evolution more efficient for problems involving phonons, it is still more computationally expensive when compared to a pure electronic model. As a result, the lesser Green's functions are only obtained for short $L_s=80$ chains with $T_s = 20[t_h^{-1}]$ with the Hubbard-extended Holstein model \cite{tangTracesElectronphonon2023}, even though more computational resources have been used when compared to the Hubbard model with $L_s=120$. All-in-all, this gives a fast Fourier transform (FFT) frequency spacing $\delta\omega \approx 0.3 t_h$ (not taking into consideration any zero-padding or broadening), which is on the same scale as the typical phonon frequency and makes resolving phonon-frequency-related spectral features a very challenging task.

To demonstrate how our method can help to resolve phonon-induced spectral features, we consider the Hubbard-extend Holstein model at half-filling with the following parameters: $U=8t_h$, $\omega_0=0.6t_h$, $g_0=0.3t_h$ and $g_1=0.15t_h$. We use tDMRG with dynamical LBO to get the simulated lesser Green's function with $L_s=80$ and $T_s=20[t_h^{-1}]$, shown in the gray area in Fig.~\ref{fig:phonon_akw}(a). We then use our method to get the predicted lesser Green's function with $L_p=160$ and $T_p=40[t_h^{-1}]$. The overall pattern in the predicted data looks consistent with the short-time data for $20[t_h^{-1}] < t \le 30[t_h^{-1}]$, and there appears to be some deviations at later times near the boundary of the light cone. These deviations will have a limited impact on the final spectra due to the use of the Gaussian window function multiplied into the raw time data prior to the Fourier transform.

In Figs.~\ref{fig:phonon_akw}(b) and (c), we show the comparison of $\mathcal{A}(k, \omega)$ obtained from the simulated $\mathcal{G}^{<}_s(r,t)$ and predicted $\mathcal{G}^{<}_p(r,t)$, respectively. The higher resolution provided by our method allows for the observation of phonon-induced spectral features not visible in the raw simulated spectra. At phonon frequency $\omega_0$ (indicated by the blue dashed line) below the Fermi energy (indicated by the black dashed line), there are 4 ``discontinuities" appearing in the holon band at points $-k_2, -k_1, k_1, k_2$  where $0 < k_1 < \pi/2 < k_2 < \pi$. 
There is also a very weak replica of the spinon band, separated by the phonon frequency $\omega_0$ from the main spinon band between $-k_1$ and $k_1$. Each of these features cannot be observed in the spectra obtained from the raw simulated data due to its reduced spectral resolution. 

\section{Discussion}
We have demonstrated a linear autoregressive machine learning-based framework to extend a time-dependent correlator to longer times and distances, using nearby space-time data obtained from a smaller simulation that captures the short-distance and short-time behavior. This results in spectra with much higher frequency and momentum resolution with virtually no additional computational cost. 

We first demonstrated the effectiveness of our method by computing $\mathcal{A}(k,\omega)$ of a 1D Hubbard model at half-filling. The resulting spectra from our method have higher quality and resolution compared to that obtained from the raw simulated data (see Fig.~\ref{fig:hubbard_akw}). It also matches the well-documented results for the one-dimensional Hubbard model in the literature \cite{senechalSpectralWeight2000, benthienSpectralFunction2004, kohnoSpectralProperties2010, chenAnomalouslyStrong2021, esslerOneDimensionalHubbard2005}. 

We then turned to a much more computationally challenging task of computing $\mathcal{A}(k,\omega)$ for the Hubbard-extended Holstein model. When compared to the pure Hubbard model, the additional phonon degrees of freedom limit simulations to shorter distances and much shorter times for reasonable computational cost, presenting a challenge to resolve phonon-related spectra features, as shown in Fig.~\ref{fig:phonon_akw}(b). Our method clearly improved the spectral resolution and allowed us to resolve the features associated with the phonon degrees of freedom and {\it el-ph} coupling, as shown in Fig.~\ref{fig:phonon_akw}(c). 


We note that there are limitations to this simple method. Firstly, we assume there is a light cone for the time-dependent correlator so that the non-zero time-dependent correlator data progressively expand from short to long distances over time. This essentially allows our method to progressively predict later time and longer distance data from the prior time and shorter distance data. This assumption holds for gapped systems, whereas gapless systems will require more sophisticated methods. Secondly, we assumed a simple linear predictor with fixed $\theta$ parameters. A more sophisticated model, perhaps including non-linearity and better modeling of space- and time-dependent parameters, could further improve the results. Nevertheless, our method can be applied easily to gapped systems for improving both the momentum and frequency resolution of spectra, especially useful when only short-distance and short-time correlator data can be obtained due to prohibitive computational cost.

\begin{acknowledgements}
This work was supported by the U.S. Department of Energy, Office of Basic Energy Sciences, Division of Materials Sciences and Engineering, under Contract No.~DE-AC02-76SF00515. The computational results utilized the resources of the National Energy Research Scientific Computing Center (NERSC), a U.S. Department of Energy, Office of Science User Facility, using NERSC award BES-ERCAP0031424. C.
Jia acknolwedges the support from Center for Molecular Magnetic Quantum Materials, an Energy Frontier Research Center
funded by the U.S. Department of Energy, Office of Science,
Basic Energy Sciences under Award no. DE-SC0019330. Some of the computing for this project was performed on the Sherlock cluster. We would like to thank Stanford University and the Stanford Research Computing Center for providing computational resources and support that contributed to these research results.
\end{acknowledgements}

\bibliography{reference}
\end{document}